\documentclass[prb,twocolumn,a4paper,floatfix,superscriptaddress,unsortedaddress,showpacs,showkeys]{revtex4-1}

\usepackage{dcolumn,amsmath,xspace}
\usepackage{graphicx}
\usepackage{multirow}

\usepackage{epstopdf}

\newcommand{\degC}{\ensuremath{~^{\circ}\text{C }}}
\newcommand{\sixrt}{\ensuremath{(6\sqrt{3}\!\times\!6\sqrt{3})\text{R}30^\circ}~}

\begin{document}

\title{Microscopic correlation between chemical and electronic states in epitaxial graphene on SiC$(000\bar{1})$}

\author{C. Mathieu}
\affiliation{IRAMIS/SPCSI/LENSIS, F-91191 Gif-sur-Yvette, France}
\affiliation{CEA,  LETI, MINATEC Campus, F-38054 Grenoble Cedex 09, France}
\author{N. Barrett}
\email[Correspondence should be addressed to ]{nick.barrett@cea.fr}
\author{J. Rault}
\author{Y.Y. Mi}
\affiliation{IRAMIS/SPCSI/LENSIS, F-91191 Gif-sur-Yvette, France}
\author{B. Zhang}

\author{W.A. de Heer}
\affiliation{The Georgia Institute of Technology, Atlanta, Georgia 30332-0430, USA}
\author{C. Berger}
\affiliation{The Georgia Institute of Technology, Atlanta, Georgia 30332-0430, USA}
\affiliation{CNRS/Institut N\'{e}el, BP166, 38042 Grenoble, France}
\author{E.H. Conrad}
\affiliation{The Georgia Institute of Technology, Atlanta, Georgia 30332-0430, USA}

\author{O. Renault}
\affiliation{CEA,  LETI, MINATEC Campus, F-38054 Grenoble Cedex 09, France}

\begin{abstract}
We present energy filtered electron emission spectromicroscopy with spatial and wave-vector resolution on few layer epitaxial graphene on SiC$(000\bar{1})$ grown by furnace annealing. Low energy electron microscopy shows that more than 80\% of the sample is covered by 2-3 graphene layers. C1s spectromicroscopy provides an independent measurement of the graphene thickness distribution map.  The work function, measured by photoelectron emission microscopy (PEEM), varies across the surface from 4.34 to 4.50eV according to both the graphene thickness and the graphene-SiC interface chemical state. At least two SiC surface chemical states (i.e., two different SiC surface structures) are present at the graphene/SiC interface. Charge transfer occurs at each graphene/SiC interface. K-space PEEM gives 3D maps of the $|\bar{k}_\parallel |$ $\pi - \pi^*$ band dispersion in micron scale regions show that the Dirac point shifts as a function of graphene thickness. Novel Bragg diffraction of the Dirac cones via the superlattice formed by the commensurately rotated graphene sheets is observed. The experiments underline the importance of lateral and spectroscopic resolution on the scale of future electronic devices in order to precisely characterize the transport properties and band alignments.
\end{abstract}
\vspace*{4ex}

\maketitle
\newpage

\section{Introduction\label{S:Intro}}
With the demonstration of GHz FETs based on epitaxial graphene grown on SiC,\cite{Moon_EDL_09,IBM} this material has become the leading candidate for graphene-based electronics. Nonetheless, exploiting the remarkable properties of graphene for carbon based electronics remains an important challenge. The band structure and transport properties of graphene must be either preserved or modified in a reproducible fashion on typical device scales. While a good deal of work has already focused on monolayer graphene grown on the SiC(0001) (Si-face),\cite{Riedl_JPhysD_10} the ability to grow thin graphene films on the SiC$(000\bar{1})$ (C-face) has only recently been demonstrated.\cite{First Cface} C-face films offer a particular advantage because of their rotational stacking that effectively decouples adjacent graphene layers.\cite{Hass_PRL_08,Sprinkle_PRL_09}  This leads to very high mobilities\cite{Berger06,Orlita_PRL_08}  and allows devices to be less sensitive to thickness variations. While these systems continue to make progress towards realistic carbon electronics, significant research problems remain.  One in particular is the study and control of the graphene-SiC interface. 
  
 When graphene is grown on the Si-terminated SiC on SiC(0001) (Si-face), the first graphene layer grows on an insulating carbon buffer layer with a \sixrt symmetry.\cite{Hass_JPhyCM_08}  This buffer layer has a graphene structure and can be isolated from the SiC by intercalating hydrogen between the SiC substrate and the buffer layer.\cite{Riedl_PRL_09}  However, mobilities in this isolated buffer layer remain low suggesting either some type of prior disorder in the layer before hydrogenation or an effect caused by the hydrogenation itself.  Very little is known about the C-terminated SiC$(000\bar{1})$ graphene-SiC interface.  However, it is known that the C-face and Si-face interfaces must be very different.\cite{Starke_JPCM_09}   X-ray studies show that the atomic density gradient at the interface is different for Si-face and C-face,\cite{Hass_PRB_08,Hass_PRB_07} and core level photoelectron spectroscopy shows clear differences between the two interfaces.\cite{Emtsev_PRB_08}  Also, unlike the Si-face, the C-face interface is known to have two coexisting structures (at least in the early growth phase).  Scanning tunneling microscopy (STM) studies have shown that, in UHV growth conditions, poorly ordered $(2\!\times\!2)$ and $(3\!\times\!3)$ surface reconstructions exist below the first graphene layer.\cite{Hiebel_PRB_08} However, these structures may disorder or simply not exist when growth occurs at higher temperatures.  In high temperature furnace growth, there is no real evidence that the interface has a reconstruction.  It is either an ordered $(1\!\times\!1)$ or a disordered reconstruction. The most important observation demonstrating that the C-face interface is very different from the Si-face is that C-face graphene has a rotational stacking very different from that of Si-face graphene even when growth temperatures are the same.\cite{Hass_JPhyCM_08,Hass_PRL_08,Sprinkle_PRL_09} This implies that the registry forces, and thus the interfaces, must be different on the two surfaces.  Finally the rapid growth of C-face graphene at temperatures lower than those observed on the Si-face point to a significant difference in the chemistry of the interface.\cite{Hass_JPhyCM_08}

In this work we present a detailed study of the C-face graphene-SiC interface. To date, most of the experimental techniques used to study the atomic and electronic structure are area-averaged and are therefore not sensitive to variations on the micron scale. Near field methods such as scanning tunneling microscopy (STM) can reduce the probed area by two orders of magnitude.  However STM does not probe the interface itself and still provides very little information on the intervening length scales, which are precisely those of interest in many potential device applications. Recently this intermediate length scale has begun to be explored using Low energy electron microscopy (LEEM). Luxmi et al.\cite{Luxmi_PRB_10} have studied the morphology of both UHV and argon furnace growth C-face graphene. Their work revealed a great deal of graphene thickness variation in these thick argon grown films.

We focus on the spatial variation of both the electronic structure and chemical bonding of C-face graphene-SiC interface. The studies were carried out on thin C-face graphene films grown by a controlled Si sublimation technique. In particular, we investigate the chemical homogeneity of the interface, and correlate it with changes in doping, the graphene work function and graphene's 2D band structure near the Fermi level. We show that the interface is very complicated with local chemical changes that are not all associated with the local graphene thickness.  To carry out these studies, we use LEEM,  photoemission electron microscopy (PEEM), and X-ray photoemission electron microscopy (X-PEEM).  In addition, by using a suitable lens configuration the focal (or diffraction) plane of the PEEM can be imaged to give parallel momentum resolved dispersion curves $E(k_x,k_y)$.  This technique is known as $k$-resolved photoemission electron microscopy ($k$-PEEM). Imaging the focal plane in PEEM produces a map for all azimuths simultaneously. Conservation of the component of the electron wave vector parallel to the sample surface automatically transforms this map into one of photoelectron intensity as a function of $(k_x,k_y)$, that is a horizontal cut in reciprocal space. Combined with energy analysis, this produces an image in reciprocal space of the local intensity as a function of wave vector parallel to the surface. For example, the Fermi surface can be acquired in a single shot experiment.\cite{Kromker_RSI_08}

\section{Experiment\label{S:Exp}}
The substrate used in these studies was a 6H conducting SiC$(000\bar{1})$ from Cree Inc. 
Before graphene growth, the sample was first $\text{H}_2$ etched for 30min at 1400\degC.
The sample was then grown in an enclosed graphite RF furnace using the confinement controlled sublimation process, CCS.\cite{First Cface}  The growth was done at 1475\degC for   20mins.

Before all measurements, the sample was annealed at 500\degC for 1 minute in UHV to remove surface contamination.  The surface cleanliness was checked by Auger electron spectroscopy and X-ray photoemission (XPS). LEEM was used to quantify the graphene layer thickness. LEEM experiments were carried out using a commercial Elmitec PEEM/LEEM III with base pressure of $4\!\times\!10^{-8}$Pa. The energy filtered X-PEEM experiments were conducted on TEMPO beam line of the SOLEIL synchrotron (Saint Aubin, France) using a NanoESCA X-PEEM (Omicron Nanotechnology GmbH).\cite{Escher_JPhyCM_05,Escher_JESRP_10}  A double pass hemispherical energy analyzer was used to compensate single analyzer aberrations. This resulted in a PEEM energy resolution of 0.2eV with a lateral resolution of $\sim\!100$nm for core level emission. Experiments were conducted in Ultra High Vacuum (UHV) at a pressure of $6\!\times\!10^{-9}$ Pa. 

The X-PEEM image series was acquired over the photoemission threshold region and the C1s core level ($h\nu$=654.3eV). A $53\mu$m filed of view (FoV) was used with 12 kV extraction voltage. For the real space PEEM mode (threshold and C1s), a contrast aperture of $70\mu$m was used, the lateral resolution was estimated to be 100 nm. The analyzer entrance slit was set to 0.5 mm with a 100eV pass energy to give a resolution of 200 meV. The resolving power of the TEMPO beamline is approximately 5000, giving an overall estimated energy resolution better than 250 meV. Dark images were acquired with the MCP turned off in order to remove camera noise. Flat Field images were acquired to correct for MCP defects. The parabolic non-isochromaticity of the instrument was corrected for all images.\cite{Escher_JESRP_10} 

The $k$-PEEM results were acquired using the same incident x-ray spot position on the sample as the X-PEEM analysis (beam size $\sim\!100\mu\text{m}\!\times\!50\mu\text{m}$ provides uniform illumination).  Because of the high extraction voltage between the sample and the lens, the wave vector resolution and the dimensions of the reciprocal space image are independent of the photon energy for the typical spectral ranges used in these experiments. The lateral spatial resolution in the $k$-PEEM mode was purposefully reduced by operating with a full open aperture.  This was required  to image a sufficient portion of reciprocal space in order to cover a full Brillouin zone. In this setup, the area of interest on the sample surface is chosen by a field aperture situated in an intermediate image plane that was closed down to about $7\mu$m.  A transfer lens then projected the $1500\mu$m diameter disk of the focal plane via the energy analyzer onto the detector giving a 2D $k$-space dimension of about $\pm 2.5\text{\AA}^{-1}$ around the $\Gamma$-point. The spectrometer resolution was 200 meV, the photon bandwidth ~20 meV, and the wave vector resolution $\sim\!0.05 \text{\AA}^{-1}$. The detector response was corrected by the flat field of the detector and camera defects were eliminated using dark  images. Operating this way gave a  $\sim\!1\mu$ lateral resolution for the band structure imaging.
 
\section{Results\label{S:Results}}
\begin{figure}
\includegraphics[width=7.5cm,clip]{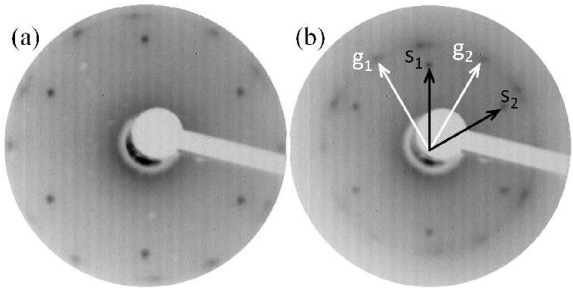}
\caption{LEED patterns from a nominally 3-layer C-face graphene sample with primary electron energy of (a) 76eV and (b) 126eV. 
} \label{F:LEED}
\end{figure}

While Si-face graphene is known to grow oriented $30^\circ$ relative to the principle SiC$\langle 21\bar{3}0\rangle$ direction, multilayer C-face graphene is known to be stacked with sheets within the stack having multiple rotation angles peaked at $30^\circ$ and $0\pm\!\sim\!7^\circ$. These C-face graphene rotations are due to interleaved rotated graphene sheets with non-Bernal stacking (i.e non-$60^\circ$ rotations). In thin C-face samples there are only a few rotations. This is demonstrated in the LEED patterns in Fig.~\ref{F:LEED}.

Because the films are thin, the LEED pattern clearly shows the six-fold SiC substrate spots ($S_1$ and $S_2$). In addition, the LEED shows a set of graphene spots rotated by $30^\circ$ with respect to the $S_2$ ($g_1$ and $g_2$) and a second set of spots rotated $\sim\!4-5^\circ$ relative to $S_2$. Weaker graphene arcs around $0^\circ$ are also visible. The SiC substrate LEED pattern is $1\!\times\!1$ and no additional diffraction spots that would suggest a significant reconstruction are visible. However, these LEED images are macroscopic area-averaged results, and are not expected to be sensitive to microscopic variations in the interface structure. More detailed spatial information is obtained using LEEM and X-PEEM as discussed below.

\subsection{LEEM\label{S:LEEM}}
\begin{figure}
\includegraphics[width=8.5cm,clip]{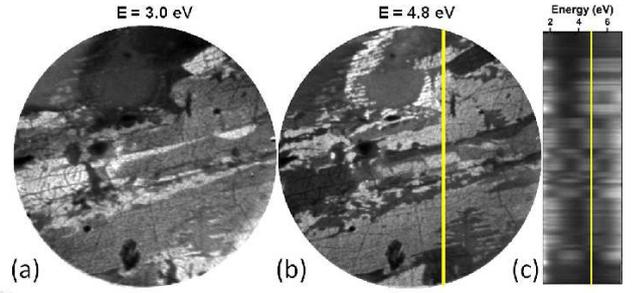}
\caption{(a)-(c) Typical LEEM images ($10\mu$m FoV) of the graphene surface as a function of electron energy (a) 3.0eV (b) 4.8eV (c) Reflectivity spectra extracted along the dotted line in (b) showing clearly the variation in the number of intensity minima and therefore the number of graphene layers.} \label{F:LEEM}
\end{figure}

LEEM data were obtained by using the (0,0) specular back-scattered electron beam. Figure \ref{F:LEEM} shows typical bright-field images with a FoV of $10\mu$m for electron energies of 3.0 and 4.8eV. A full image series was acquired by varying $E$ from 1.5 to 12.9eV (using a 50meV step size). The low energy onset of the back-scattered electron signal depends on the  potential just above the surface and thus on the local work function.  Clear differences in contrast as a function of $E$ are observed across the field of view suggesting distinct work function for different regions of the surface.

\begin{figure}
\includegraphics[width=6.5cm,clip]{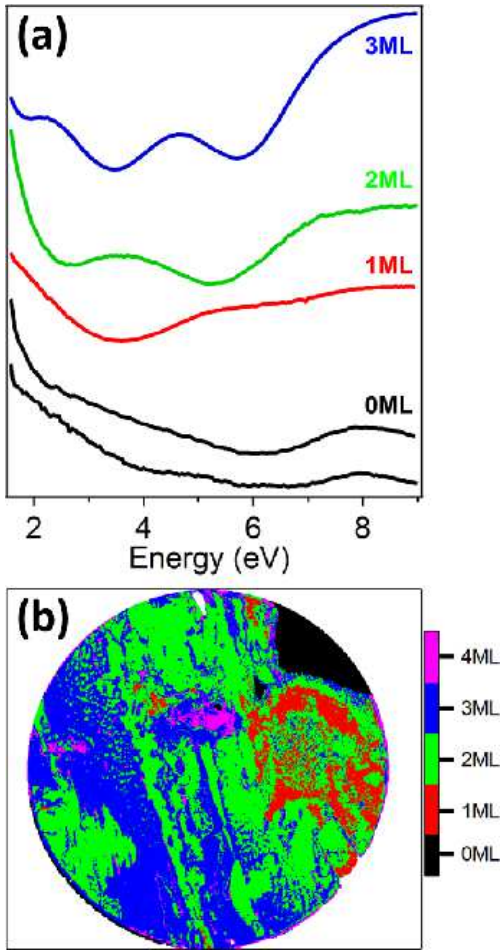}
\caption{(a) Typical reflectivity curves extracted from the LEEM dataset showing 0 to 3 oscillations below 7.5eV.  (b) Graphene thickness map (FoV=$10\mu$m) generated by counting the number of minima in the reflectivity curves. } \label{F:LEEM_Osc}
\end{figure}

In two-dimensional layered systems there are oscillations in the LEEM reflectivity at low electron energies.\cite{Hibino_PRB_08,Ohta_NJP_08,Chung_PRL_03}  Several groups have used these reflectivity oscillations to determine the number of graphene layers in the epitaxial film. Figure \ref{F:LEEM}(c) shows the spectra along the dotted line in Fig.~\ref{F:LEEM}(b). We observe oscillations between 1.5 and about 7.5eV. In graphite there are band gaps below 0eV and above 7eV along the $\Gamma A$ direction. In an ideal multilayer graphene film, the number of minima in the reflectivity between successive Bragg peaks gives directly the number of graphene layers. We have similarly extracted pixel-by-pixel reflectivity curves from the image stack and mapped the LEEM reflectivity across the $10\mu$m FoV and used the number of minima in each reflectivity curve to produce a thickness map of the C-face graphene film. Figure~\ref{F:LEEM_Osc}(a) shows the typical reflectivity curves for each distinct contrast region in the FoV.  Two slightly different curves without a clear oscillation between 1.5 and 7eV are observed. They are both attributed to the zeroÕth layer or C-terminated layer of the SiC substrate. This layer is thought to interact strongly with the SiC substrate through $\text{sp}^3$ bonding. Luxmi et al.\cite{Luxmi_PRB_10} have shown even flatter reflectivity curves for the zeroÕth layer (0ML) for surfaces prepared at higher temperature under an argon back pressure. The thickness map constructed from the intensity minima is shown in Fig.~\ref{F:LEEM_Osc}(b) using the same color coding. The data shows that more than 80\% of the surface is covered by 2- or 3-layer graphene.

The reflectivity oscillation can be understood as quantum interference between electrons reflected by different graphene layers.\cite{Hibino_PRB_08} Hibino et al.\cite{Hibino_PRB_08} pointed out that although the conduction band is continuous in bulk graphite between 4.3 and 11eV along $\Gamma A$, few layer graphene should have discrete states, thus the reflectivity oscillations are correlated with the electron structure of the thin films. The quantized conduction band states enhances the transmission of the incident electrons producing the dips in the reflectivity curves. This is confirmed by the good agreement between the experimental minima positions and the resonant energies predicted by a tight-binding calculation.   For an $m$-layer thin film the bulk band dispersion has discrete energy states when the wave vector satisfies the quantized condition $E=\epsilon-2t\cos{K\alpha}$, where $\epsilon$ is the energy of the band center, $t$ is the transfer integral, and $\alpha$ is the interlayer distance. At these values, the dips in the reflectivity are predicted by projecting the wave vector onto the calculated band structure along $\Gamma\!-\!A$ direction, the normal to the graphene layers.
 
 \begin{figure}
\includegraphics[width=7.0cm,clip]{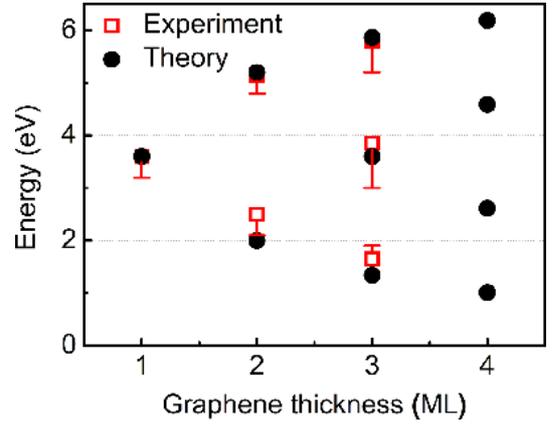}
\caption{The position (boxes) of dips in reflectivity curves from Fig.~\ref{F:LEEM}(c) and Fig.~\ref{F:LEEM_Osc}(a), compared with theoretical tight binding estimation based on tight-binding model ($\bullet$). The error bar on the experimental data indicates the correlations between the spread in the energy of the reflectivity minima. } \label{F:LEEM_STATES}
\end{figure}

Figure~\ref{F:LEEM_STATES} shows the comparison of experimental reflectivity minima with the discrete energy levels predicted by the tight-binding theory. The spread in the experimental energy of the reflectivity minima in different regions of the FoV is indicated as an error bar. The centre energy of the tight-binding theory calculation is at 3.6eV compared to 3eV used by Hibino for Si-face graphene.\cite{Hibino_PRB_08} Note that the third calculated minimum (highest energy) in the nominally 3ML region is at a slightly higher energy whereas the first and second minima are at slightly lower energy than the experimental data. This trend is also seen in the 2ML minima.  The difference may be due to the approximations inherent in the tight binding calculation.  

\begin{figure*}
\includegraphics[width=16.0cm,clip]{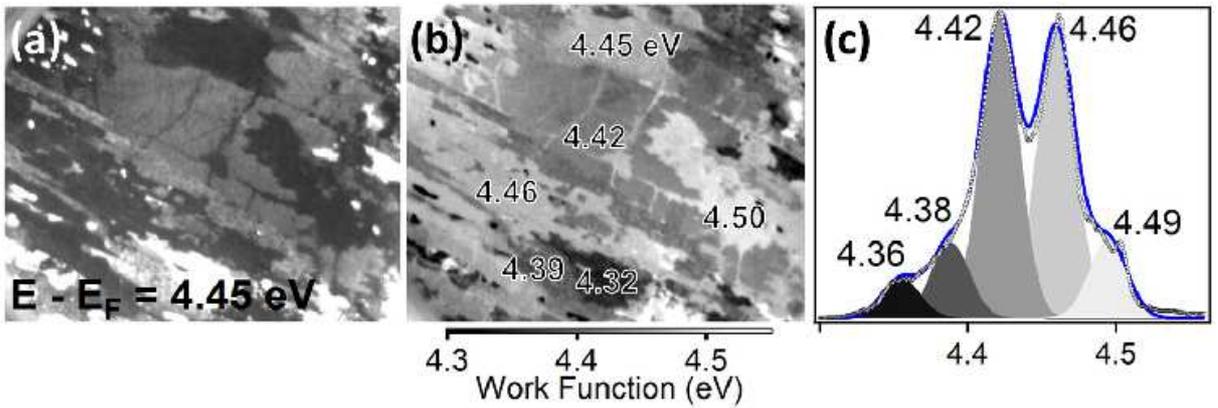}
\caption{(a) Threshold image at $E-E_F$=4.45eV with a $53\mu$m field of view showing clear intensity contrast due to work function variation across the surface. (b) Work function map of the FoV obtained from a pixel by pixel analysis of the threshold spectra. (c) Histogram of the work function values in (b) showing five distinct Gaussian distributions, corresponding to five distinct work function values.} \label{F:Threshold}
\end{figure*}

\subsection{X-PEEM\label{S:XPEEM}}
While LEEM measurements give information on the structural spatial variation of the graphene thickness, lateral variations in the electronic structure of epitaxial graphene have never been mapped. XPEEM offers a unique method to begin to understand the role the SiC-graphene interface plays in grapheneÕs electronic structure. In this section we present the first sub-micron chemical and electronic mapping of the graphene-SiC interface. 

\subsubsection{Work function\label{S:WF}}
Probing the transition from the mirror reflection of the electrons to the back scattering regime, commonly referred to as mirror electron microscopy to low energy electron microscopy (MEM-LEEM) transition is highly sensitive to the local variations of the electric potential just above the surface, as small differences in the latter determine large differences in the electron reflectivity. In energy-filtered PEEM, at high photon energies (654.3eV in these experiments), the photo-emitted intensity at threshold is directly related to the work function. An example of a raw image obtained in the threshold region is shown in Fig.~\ref{F:Threshold}(a). As $E-E_F$ is scanned, dark areas become bright and vice-versa, giving rise to spatial contrast as a function of photoelectron kinetic energy. This is direct evidence for a distribution of work function values over the sample surface.
 
After correction for the Schottky effect due to the high extractor field, $\Delta E\!=\!98$ meV for 12 kV,\cite{RBRH06} the photoemission threshold spectra can be used to directly measure the local work function. The threshold spectrum is extracted from each pixel in the FoV (pixel area $65\times65 \text{nm}^2$). The position of the threshold is obtained from a fit using a complementary error function,
 
\begin{equation}
I(E)= A\cdot\text{erfc}  \lbrack\frac{\Phi_o-E}{\sqrt{2}\sigma}\rbrack+ I_\text{min}. \label{E:I(E)}
\end{equation} 
where $\Phi_o$  is the work function and $\sigma$ the half-width of the rising side of the secondary electron peak (0.1eV here). We note that because the theoretical shape of the photoemission onset is modeled, this method is a more reliable way of obtaining absolute work function values than simply extrapolating a straight line down to zero intensity. The results of this analysis are presented in the form of a map of the work function within the field of view in Fig.~\ref{F:Threshold}(b). A histogram of the work function values across the whole FoV is shown in Figure~\ref{F:Threshold}(c). Five Gaussians, with width $\pm 25$meV, are able to describe the work function frequency distribution suggesting that there are only five distinct values of the work function. Actually, as we show when we discuss the XPEEM results, the peak in the work function distribution centered at 4.46eV includes two slightly different different work functions ($\Delta\Phi\!=\!10$meV). While these two regions of the surface have nearly the same work function they can be distinguished by their very different core level spectra. 

The work function distribution spans a range from 4.34eV to 4.50eV; a range well below the work function of bulk graphite (4.6eV); thus our results confirm that no significant part of the surface within the FoV consists of many layer graphene or graphite. On the contrary, this epitaxial film is indeed near-uniform few layer graphene, with local variations in the graphene coverage. However, we cannot necessarily attribute each peak in the work function distribution to a distinct graphene thickness. For example, recent Kelvin force microscopy showed that the work function difference between one and two layer graphene is 0.135eV.\cite{Filleter_APL_08}  The graphene thickness calibration, based on the local C1s XPEEM spectra presented below, excludes such a difference between regions with 1 and 2 layer graphene, therefore the observed variations cannot be due to only thickness changes.  This is particularly true for regions of the surface with work function values of 4.46-4.49eV.  For these regions the C1s spectra (as discussed in the next section) indicate that the graphene film is in fact very thin.

\begin{figure}
\includegraphics[width=8.0cm,clip]{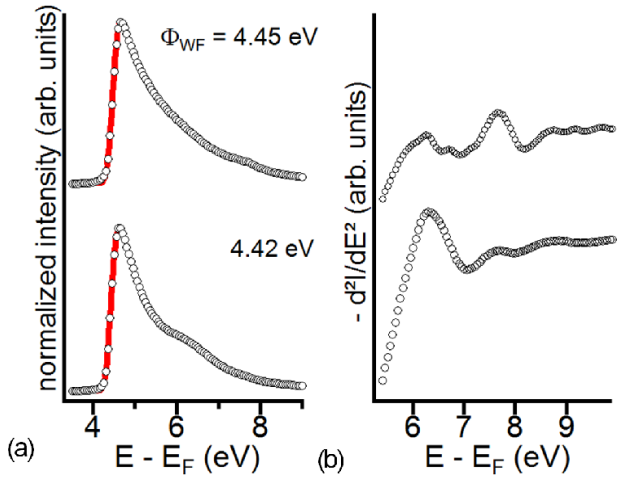}
\caption{(a) Local threshold spectra extracted from regions in Fig.~\ref{F:Threshold}(b) with work functions of 4.42 and 4.45eV. The best complementary error function fits of the rising edge of the photoemission threshold are indicated by the solid line. (b) Second derivative smoothed threshold spectra from (a) showing peaks corresponding to the empty conduction band structure.} \label{F:Thres_2ndDiv}
\end{figure}

There are also more substantial changes to the threshold spectra than the work function measurements suggest.  To demonstrate this, Fig.~\ref{F:Thres_2ndDiv}(a) shows two examples of threshold spectra for regions of the sample with a small work function difference of  $\sim\!30$meV (4.42 and 4.45eV).  These regions will be analyzed in more detail using both core level XPEEM and the $k$-PEEM data in the next section. There are interesting variations in the structure of the secondary electron (SE) peak up to 4-5eV above the vacuum level (i.e. above the photoemission threshold). The SE peak structure can be related to the conduction band.\cite{Feder_SSC_78}   Figure \ref{F:Thres_2ndDiv}(b) shows the negative of the second derivative threshold spectra. The data have been smoothed in order to more clearly see the peak structure. The low work function regions in Fig.~\ref{F:Threshold}(c) have a single main structure in the SE at $\sim\!6.0$eV whereas the higher work function regions have a clear double structure at $\sim\!6.0$eV and $\sim\!7.50$eV. The structure around 6eV could be due to the bulk SiC bands observed along $\Gamma A$, that extend from 5.6 to above 6eV.\cite{Persson_JAP_97}  Other conduction bands also disperse along the bulk directions $ML$ and $HK$. In fact, all the high work function spectra show a double peak structure in the SE whereas the low work function spectra show only one peak. This sort of SE structure has already been observed in threshold XPEEM analysis of graphene on SiC(0001).\cite{Hibino_PRB_09} It has been known for a long time that bulk graphite produces an intense secondary electron signal at 7.5eV above the Fermi level (about 3eV above the vacuum).\cite{Willis_PRB_71}  For example, the photocurrent carried by the Bloch constituent of the time-reversed LEED wave-function is indeed a maximum near 7.5eV.\cite{Barrett_PRB_05} Thus although the work function difference in Fig.~\ref{F:Threshold}(a) is small ($\sim\!30$meV), we can identify the presence of a band at 7.5eV, which for thicker samples, could develop into the typical structure of graphite. 

\subsubsection{C1s core level\label{S:C1s}}

\begin{figure}
\includegraphics[width=8.5cm,clip]{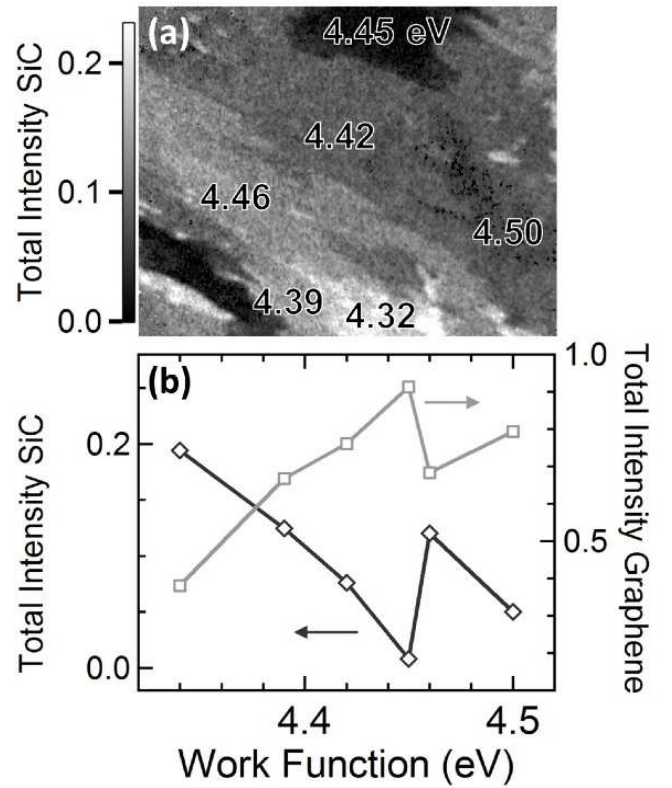}
\caption{(a) Intensity map (arbitrary units) of the SiC substrate component of C1s spectrum after background subtraction: the darker regions correspond to lower intensity and therefore thicker graphene, FoV$53\mu$. (b) the C1s intensities from graphene (squares) and from SiC (diamonds) as a function of work function}. \label{F:C1s_map}
\end{figure}

A series of energy filtered images have been acquired over the full C1s spectra in the same FoV as the threshold data, allowing a pixel-by-pixel extraction of the local C1s spectra that can then be directly compared to the work function map. The C1s spectrum has two main components, one assigned to the SiC substrate near 283eV, and the other to graphene near 285eV.\cite{Emtsev_PRB_08} Figure~\ref{F:C1s_map}(a) shows a map of the total area of the SiC component.  Because thicker graphene regions attenuates the photoelectrons from the SiC more, the map in Fig.~\ref{F:C1s_map}(a) is a good estimate of the variations in the graphene thickness within the FoV. Comparison of the C1s intensity map with the work function map in Fig~\ref{F:Threshold}(b) shows that there is no simple one-to-one correlation between graphene thickness and work function.  Instead we will show that much of the contrast variations are a combination of both work function and core level spatial variations, indicating a complex chemical structure in the SiC-graphene interface.

\begin{figure}
\includegraphics[width=7.0cm,clip]{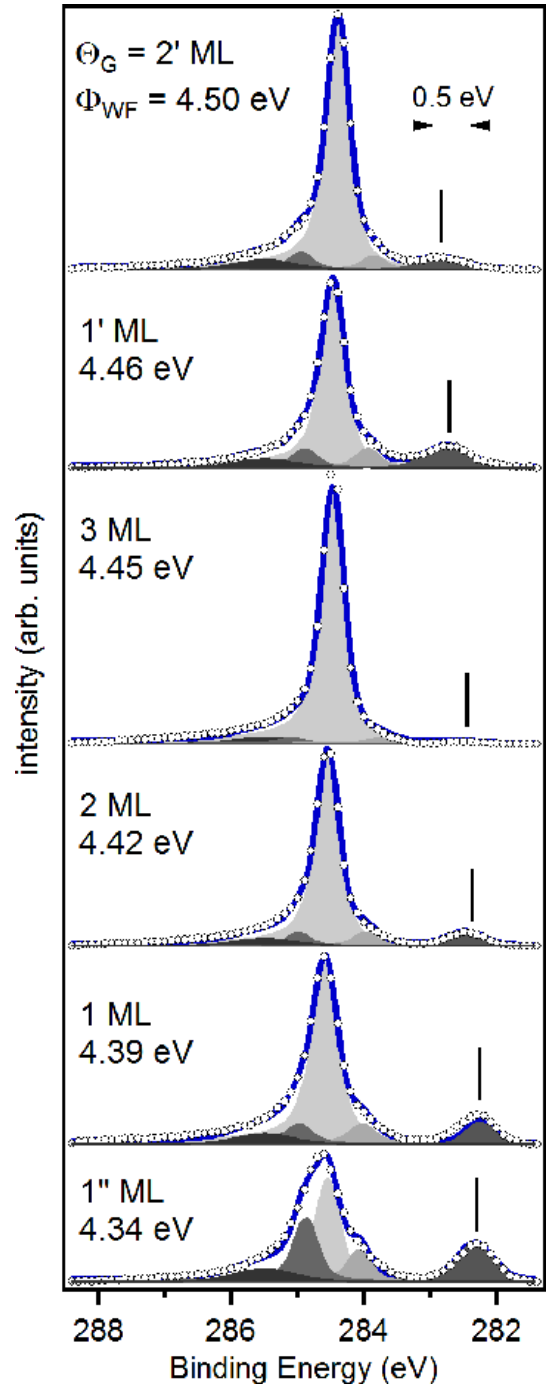}
\caption{C1s core level spectra extracted from the regions shown in Fig.~\ref{F:C1s_map} together with the best 5-peak fits. The vertical lines indicate the position of the peak attributed to the SiC substrate. The main graphene peak is light grey, the contamination peak is in black, the SiC component is dark grey. The two other graphene peaks, labeled HBE-G and LBE-G, flank the main graphene peak. The thickness (calibration is described in the text) and work function are given for each spectra.} \label{F:C1s_spec}
\end{figure}

In order to obtain more chemical-spatial detail, local, C1s core level spectra were extracted from each of the regions identified in Fig.~\ref{F:C1s_map} and are presented in Fig.~\ref{F:C1s_spec}. The graphene and SiC intensities show significant variations from one region to another, not only in total intensity but also in terms of the fine structure of both the graphene and SiC components. To fit the spectra, we use five peaks, after subtraction of a Shirley background (a linear background fit was also tested but does not significantly change the results). Three components are necessary to fit the main peak of the spectrum that represents the graphene. For each of these components a Doniach-Sunjic lineshape was used with a 0.2eV Lorentzian width, 0.3eV Gaussian width and 0.05 asymmetry factor. A one component Gaussian lineshape (FWHM of 0.5eV) is used to fit the low binding energy peak ascribed to the SiC substrate. A Gaussian is more suitable for a wide gap semiconductor. The binding energy of this component can vary by up to 0.5eV. Finally, a small broad component, which is always present in the C1s spectrum around 285.7eV, is ascribed to slight surface contamination due to residual gas in the vacuum chamber and fitted using a Gaussian with a FHWM = 1.1eV. The contamination component increases slightly for thinner graphene, and is largest in region with a work function of 4.34eV. The best fits are also reported in  Fig.~\ref{F:C1s_spec}.  Each spectra is correlated with a region of the sample that is specified by its local work function and thickness (determined by it C1s as described below).

The main graphene component lies between 284.30 and 284.56eV and is always the dominant contribution to the spectrum, confirming that there is graphene or a graphene like layer over the whole film. This statement is supported by the behavior of the SiC component, at lower binding energy. It is always much smaller than the main graphene component suggesting a near continuous graphene coverage. As the graphene intensity increases (i.e. the number of graphene layers, increases), the intensity of the SiC decreases. We can estimate the local graphene thickness from the relative SiC and graphene C1s intensity. Using a graphene interlayer spacing of 0.34 and a 1 nm\cite{Tanuma_SIA_91} electron mean free path for 654.3eV photons in graphite, the estimated attenuation of the substrate C1s signal is 22\% per graphene layer. Assuming a C atom surface densities for graphene ($3.8\times10^{15} \text{cm}^{-2}$) and SiC ($1.22\times10^{15}\text{cm}^{-2}$), the thickness of the graphene within the FoV is between 1 and 3 ML. This estimate is in agreement with the spread in the number of ML deduced from oscillations in the LEEM backscattered reflectivity curves.  The thickness values obtained by this method are given in Fig.~\ref{F:C1s_spec}. The work function, C1s core level binding energies and graphene thicknesses are summarized in Table \ref{tab:C1s_par}.

\begin{table}
\caption{\label{tab:C1s_par} C1s BE for the main graphene peak and for the SiC peak together, the G/SiC core level intensity ratio for different graphene thicknesses and work functions.}
\begin{ruledtabular}
\begin{tabular}{ccccr}
\multirow{1}{*}{Coverage} &\multirow{2}{*}{$\Phi_{WF}$ (eV)} &\multicolumn{2}{c}{\underline{C1s BE (eV)}}&\multirow{1}{*}{Ratio}  \\
(ML) &    &    Graphene &  Substrate & $I_G/I_{SiC}$ \\
\hline       3 & 4.45 &	284.45 &  282.52 &  112.3    \\
	       2 & 4.42 &	284.55 &  282.46 & 10.0     \\
	       2' & 4.50 &	284.30 &	282.84 &  15.8    \\
	       1 & 4.39 &	284.56 &  282.32 &   5.3     \\
	       1' & 4.46 &	284.45 &	282.75 &   5.7	     \\
	      1" & 4.34 &	284.54 &  282.32 &   2.0	   \\
 \end{tabular}
\end{ruledtabular}
\end{table}

 It is immediately obvious that there is not a simple one to one correspondence between work function and graphene coverage. The same graphene thickness is obtained in regions with significantly different work functions. We can group the six different contrast regions in
Figs. ~\ref {F:Threshold}(b) and \ref{F:C1s_map}(a) into three ÒfamiliesÓ. The first consists of graphene with 1, 2 or 3 layers. The second, denoted by prime superscript, has either 1Õ or 2Õ ML graphene, while the third, denoted by double prime has only 1ÕÕ ML graphene. 

\begin{figure}
\includegraphics[width=6.5cm,clip]{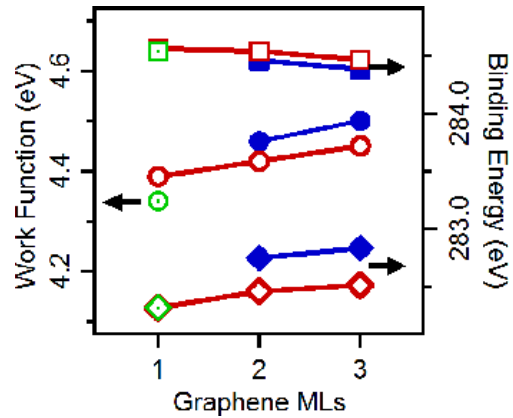}
\caption{C1s graphene BE (squares), C1s SiC BE (diamonds), work function (circles), as a function of graphene thickness. The unfilled symbols are for 1, 2, and 3 ML regions, filled symbols are for 1' and 2' ML regions. Open symbols with dots are for the 1"ML.} \label{F:BEvsN}
\end{figure}

The complexity of the graphene-SiC interface is revealed by a detailed look at the C1s spectra in Fig.~\ref{F:C1s_spec}. There are always two other graphene like peaks besides the main peak in all regions, a high binding energy (HBE-G) peak around 284.9eV and a low binding energy (LBE-G) peak at 284.0. The intensity of these two peaks is lowest for 3 ML graphene and highest for the single layer graphene, whatever the family. We therefore associate these two structures with carbon \emph{below} the graphene layers, either in the C terminating SiC layer or from some of the graphene at the SiC interface. Note that these HBE- and LBE-G peak intensities are not the same for the three single layer graphene regions. They are most intense for the 1" ML region, which also had the strongest SE structure associated with the SiC substrate. The 1 ML and 1" ML graphene have a similar C1s binding energies ($\Delta BE\!=\!100$meV). The total work function variation over the three single layer graphene regions is 120meV. Similarly, 2 ML graphene has significant differences in its C1s binding energy and in its work function. The main graphene C1s peak BE in the 2' ML graphene is shifted 250meV lower than the 2 ML graphene.  At the same time, the SiC component shifts 380 meV to higher BE in the 2' ML. However, the work function difference between the 2 ML and 2' ML regions is 80 meV. The work function, C1s graphene, and SiC binding energies are plotted together in Fig.~\ref{F:BEvsN}.

It follows that a uniform charge transfer over the whole graphene/SiC interface cannot be explained by these experimental observations. Charge transfer from the substrate to the graphene should result in a rigid shift of the electronic levels and the work function. However, in regions with the same number of graphene layers we observe a spread in the work function. The work function difference between 1 ML and 1" ML might be ascribed to the slightly higher surface contamination, but this cannot explain the work function and C1s BE difference with the 1' ML region which should have the same surface contamination. The most direct evidence for a non-uniform G/SiC interface is the shift of up to 220 meV in C1s SiC binding energy for different coverages. The trend in the SiC component binding energy emission is shown in Fig.~\ref{F:BEvsN}. If we assume that the SiC C1s signal is dominated by the first substrate layer, the trend suggests two possible interfaces. Indeed, STM studies have shown that for very thin UHV grown graphene layers there are two different interface reconstructions  ($2\!\times\!2$ and $3\!\times\!3$).\cite{Hiebel_PRB_08}  The BE variation of the C1s components as a function of thickness are also shown in Fig.~\ref{F:BEvsN}. Different core level binding energies for the same graphene thickness further supports the interpretation of a non-uniform G/SiC interface. Comparing the main G and SiC components, we see that as the SiC BE increases (charge transfer) there is a corresponding decrease in the G BE. Thus, the magnitude of the charge transfer depends on the local interface chemistry or structure. 

\subsection{$k$-PEEM Results\label{S:k-PEEM}}
In addition to the spatially resolved core level data, we are able to create local iso-intensity surfaces in $(k_x,k_y,E)$ space. In this way, we can immediately visualize the full band dispersion in all directions parallel to the graphene planes for a selected micron scale region. The objective is to correlate the chemical and electronic states obtained in XPEEM with a quantitative analysis of the band structure of the same micron scale region. Figure \ref{F:C1s_map}(a) shows a complete three dimensional dataset of the band dispersion near the Fermi level of the 2 ML region. The image series were taken from 2.9eV below the Fermi level to 0.3eV above $E_F$ in 50 meV steps, and were repeated several times in order to improve statistics without introducing camera noise. 

\begin{figure}
\includegraphics[width=6.5cm,clip]{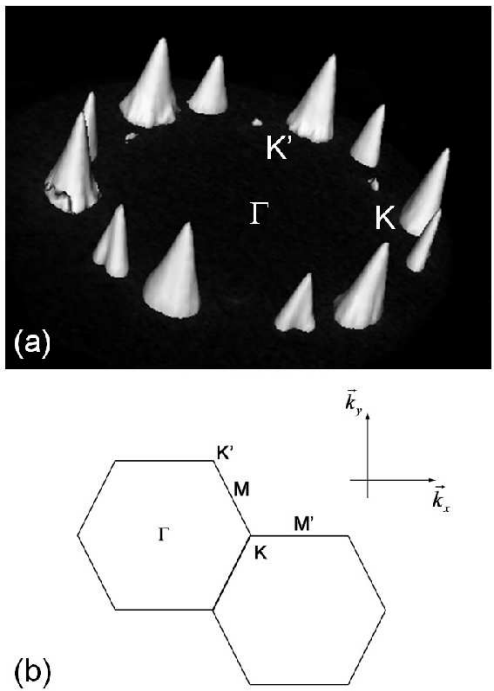}
\caption{(a) Experimental $I(k_x,k_y,E)$ data collected in in the 2 ML region of Fig.~\ref{F:C1s_map}. The principal and secondary Dirac cones at the $K$ and $K'$ points of the first Brillouin zone are clearly visible. The secondary cones are rotated by $21.9^\circ$ with respect to the primary cones. Also evident is the already documented suppression of the photoemission intensity outside the first Brillouin zone due to the pseudo-spin structure when using p-polarized light.  The tertiary cones are just visible inside the double hexagon defined by the principal and secondary cones. (b) Schematic of the Brillouin zone of graphene showing the two high symmetry directions parallel and perpendicular to $\Gamma KM'$ ($k_x$ and $k_y$, respectively) extracted from the $k$-PEEM datasets. An open source volume viewer, designed for medical imaging, was used to produce the image.\cite{program} } \label{F:E_cuts}
\end{figure}

Instead of the single set of six Dirac cones usually reported for Si-face films,\cite{Bostwick_NP_07} Fig.~\ref{F:E_cuts}(a) shows three sets of Dirac cones, which we will call principal, secondary and tertiary cones. The first two are much more intense than the latter. All three have the typical six-fold symmetry, although the tertiary cones appear inside the primary and secondary cone radius (i.e, at position $< \Gamma K$). Also note the symmetry of the primary and secondary cones. They are not full circles because of the suppression of intensity in the second zone due to matrix element effects (an effect well known in graphene).\cite{Shirley_PRB_95,Bostwick_NP_07} However, the tertiary cones have their symmetry flipped $180^\circ$. As we will demonstrate below, these three sets of cones are all due to two commensurately rotated graphene sheets.

\begin{figure}
\includegraphics[width=7cm,clip]{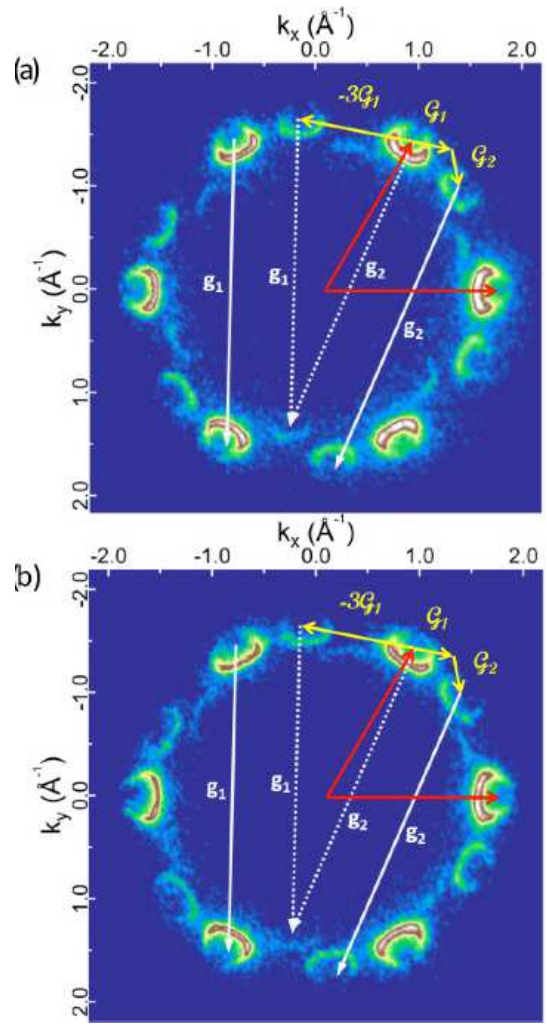}
\caption{Horizontal slice in the 3D $k$-PEEM dataset of (a) 2 ML and (b) 3 ML at a binding energy of 1.3 eV. Secondary Dirac cones (graphene reciprocal lattice constant $g_1$) are rotated by $21.9^\circ$ with respect to the principal cones (graphene reciprocal lattice constant $g_2$). The tertiary cones are obtained by diffraction of either the secondary cone by the principal lattice vector $g_1$ (dotted line), or by the primary cone by the secondary lattice vector $g_2$ (dotted line). The tertiary cones are stronger in 2 ML graphene.} \label{F:Fermi_surf}
\end{figure}

The secondary Dirac cones have reciprocal lattice vectors $\overline{\Gamma K}$ rotated by $21.9^\circ$ with respect to the primary cones. This is very close to the value of $21.8^\circ$ expected for a particular pair of commensurate rotated graphene sheets.\cite{Mele_PRB_10}  At first sight, one would be tempted to interpret the tertiary cones in terms of replicas, like those observed on the SiC(0001) Si-face.\cite{Nakatsuji_PRB_10} However, those replicas are due to registry between the \sixrt  reconstructed substrate and the graphene overlayer. On SiC$(000\bar{1})$ the graphene layers are known to form commensurate rotated layers where the average rotation between pairs is $30\pm\!\sim\!7^\circ$.\cite{Sprinkle_PRL_09} The tertiary cones in Fig.~\ref{F:Fermi_surf} are instead due to a diffraction effect caused by the supercell formed by the commensurate rotations.  To understand this, note that the supercell formed by the stacked rotated sheets is defined by reciprocal lattice vectors $\vec{\mathcal G}_1$ and $\vec{\mathcal G}_2$.  These supercell vectors are, in turn, an integer sum of the primary graphene reciprocal vectors, $\vec{g}_{1a}$, $\vec{g}_{1b}$, $\vec{g}_{2a}$, and $\vec{g}_{2b}$, from the two layers (i.e., $\vec{\mathcal G}_1\!=\!p\vec{g}_{1a}+q\vec{g}_{1b}$ and  $\vec{\mathcal G}_2\!=\!p'\vec{g}_{2a}+q'\vec{g}_{2b}$ where $p$, $q$, $p'$ and $q'$ are integers).\cite{Mele_PRB_10}

In Fig.~\ref{F:Fermi_surf}, we show two examples where two of the secondary cones are separated by linear sums of lattice vectors $\vec{\mathcal G}_1$ and $\vec{\mathcal G}_2$, i.e.,  $-3\vec{\mathcal G}_1$ and  $-(\vec{\mathcal G}_1+\vec{\mathcal G}_2)$.  Note that  $(\vec{\mathcal G}_1+\vec{\mathcal G}_2)\!=\!(\vec{g}_{1a}-\vec{g}_{2a})$  indicating that the two rotated graphene sheets are indeed a commensurate rotated pair.  For the purpose of discussion, we use the notation that $\vec{g}_{1}$'s are the graphene lattice vectors of the primary cones and $\vec{g}_{2}$'s are the graphene lattice vectors of the secondary cones.  Using this notation, it is easy to show that the tertiary cones are diffracted replicas of the primary and secondary cones.  Figure \ref{F:Fermi_surf} shows that a tertiary cone is formed by diffracting a primary cone by $\vec{g}_2$ or a secondary cone by $\vec{g}_1$. This diffraction process explains why the tertiary cones are rotated by $180^\circ$ relative to the primary and secondary cones; they are simply translated Dirac cones from a $K$ and $K'$ point of one lattice by the $\vec{g}$ of the second lattice.  Because any tertiary cones can be formed by diffraction of either of the rotated graphene lattices (primary or secondary), the two rotated graphene sheets must be stacked on top of each other with a commensurate rotation.  Separated rotated sheets cannot give rise to this type of diffraction.

\begin{figure}
\includegraphics[width=9cm,clip]{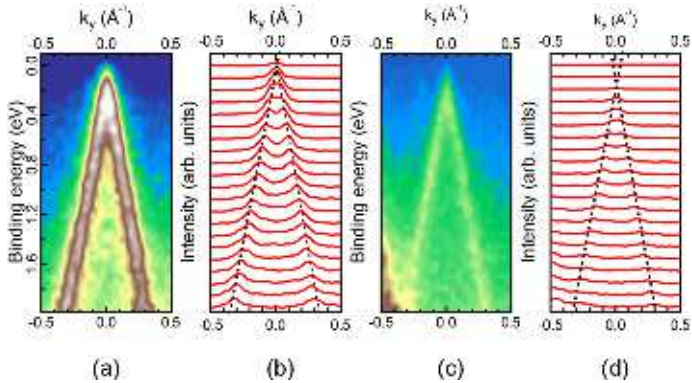}
\caption{Band dispersion as a function of  $k_y$ around the $K$-point at the first Brillouin zone boundary observed in the (a) 2ML and (c) 3ML regions. The respective momentum distribution curves over the same energy range are shown in (b) and (d). The peak positions are highlighted by the dotted lines showing the variation in the position of the Dirac point in the different regions.} \label{F:Band_y}
\end{figure}

The $\pi$-bands of the primary cones cross at a Dirac point of 0.30eV for the 2ML film and 0.25eV for the 3ML film relative to the Fermi level. Figure \ref{F:Band_y} shows cuts perpendicular to $\Gamma KMÕ$ of the primary and secondary cones for 3 ML. The linear dispersion of the $\pi$-band near the Dirac point is evident. Panels (a) and (c) show the raw experimental data around the $K$-point of the Dirac cones.  Panels (b) and (d) show the corresponding momentum distribution curves (MDCs) extracted from the data. The Fermi level calibration was checked by extracting the local spectra from a small zone around the Dirac cones. This has the advantage of minimizing noise and thus allowing a more precise location of the Fermi level. The dotted lines show that the $\pi$-bands cross at a Dirac point 50-75meV closer to the Fermi level for the principal cones. The group velocity of the quasi-particle is $1.0-1.1\!\times\!10^6\text{ms}^{-1}$ in both 2ML and 3ML regions.

This confirms that the electron doping is a function of film thickness and is consistent with the results  graphene  on the Si-face where the charge transfer induced shift of the band structure is greater for thinner films.\cite{Ohta_PRL_07}

\section{Conclusion} 
We have studied spatial variations in and correlations between the work function, chemical and electronic states of few layer graphene grown epitaxially on SiC$(000\bar{1})$. These experiments demonstrate that high thickness uniformity graphene films can be prepared using the CCS growth method. In this sample more than 80\% of the surface is 2-3 layers thick.  More importantly, they also show that while LEEM can be used to give local information on the graphene films thickness, LEEM contrast variation on C-face graphene films are not simply due to film thickness alone. X-PEEM's spatial resolution has proven to be very useful in illuminating the origin of these contrast variations. The $0.1\mu$m spatial resolution in the present experiment is much better than that currently obtained in area averaged high resolution ARPES with synchrotron radiation. The work function variations in Fig.~\ref{F:Threshold}, derived from the X-PEEM, combined with the local C1s spectra in Fig~\ref{F:C1s_spec} show an interface chemistry that is more complex than originally suspected.  

\begin{figure}[!h]
\includegraphics[width=6.0cm,clip]{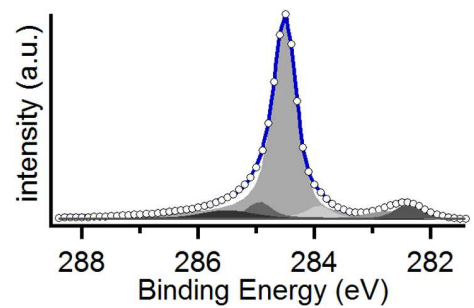}
\caption{ Area averaged C1s core level spectrum.  The average is over the $53\mu$m FoV}\label{F:LAST}
\end{figure}

To demonstrate this, Fig.~\ref{F:LAST} shows an area averaged C1s spectrum from the $53\mu$m FoV. The overall spectrum is broader and the fine structure visible in the local spectra of Fig.~\ref{F:C1s_spec} is smeared out in the area averaged spectrum. Without the spatial resolution, one would have concluded that only one broad substrate peak was present rather than a narrower peak whose energy position shifts depending on the graphene thickness. Similarly, the area averaged work function masked the complexity of charge transfer at the graphene-SiC interface.

While the X-PEEM data agrees with the general trend observed by others that the work function trends to higher values with thicker graphene films, it shows that the situation is more complicated.  For a given thickness, both the local work function and C1s BE vary appreciable. The data suggests that at least two SiC terminations with different local bonding appear to be present at the interface. We note that the long range order of these different bonding areas is not very high since the micro LEED never sees any superlattice diffraction spots. For a given graphene-SiC interface, the core level data can be interpreted within the framework of a charge transfer model between the substrate and the graphene that rigidly shifts the C1s core level. The effect of the charge transfer on binding energy is greater for carbon closer to the interface, i.e. for carbon in thinner graphene films. The correlation between the work function and the core level BE strongly supports this interpretation. On the basis of these results we suggest that there are two distinct interfacial chemical regions.

It is possible that these two regions are similar to the $2\!\times\!2$ and $3\!\times\!3$ observed by STM during the early UHV growth of C-face graphene.\cite{Hiebel_PRB_08} In the furnace grown graphene used in this results, the higher growth temperature (1550°C versus 1100°C) prevents any significant interface ordering so that no LEED spots are observed. However, Hiebel et al\cite{Hiebel_PRB_08} suggested that there is a stronger, localized interaction between the $2\!\times\!2$ reconstructed C layer and the SiC. Thus, one would expect a higher density of $sp^3$ C-Si bonding between the C termination layer and the underlying SiC. From simple charge transfer arguments, this would translate into a lower C1s binding energy. On this basis the C1s spectra of the 1'-2' ML regions could be tentatively associated with a local $2\!\times\!2$ surface.  In that case regions labeled as 1, 2 and 3 ML in Table \ref{tab:C1s_par}  would be associated with a local $3\!\times\!3$ interface.  However, it is also possible that these different regions of the surface may be due to local silicon concentrations.  This can only be resolved by performing Si 2p mapping (experiments that are now being planned).

We have also demonstrated that $k$-PEEM gives detailed band structure information from commensurately rotated graphene sheets. The easy switching between real and $k$-space imaging modes allows, within the limits of the spatial resolution and field aperture size, correlation between the chemical and electronic states of the surface and graphene-SiC interface. In this study we conclusively show that commensurately rotated graphene pairs exist.

\begin{acknowledgments}
C. Mathieu benefited from a grant of the Nanosciences programme of the French Atomic and Alternative Energy Authority. J. Rault is funded by a CEA PhD grant ÒContrat de Formation par la rechercheÓ. The work was supported by the French National Research Agency (ANR) through the ÒRecherche Technologique de Base (RTB)Ó program. We acknowledge SOLEIL for provision of synchrotron radiation facilities and we would like to thank F. Sirotti, the TEMPO beamline staff, D. Martinotti, B. Delomez and F. Merlet for technical assistance. E. Conrad, W.A. de Heer and C. Berger would like to acknowledge supported from the W.M. Keck Foundation, the Partner University Fund from the Embassy of France and the NSF under Grant No. DMR-0820382.  E. Conrad would also like to acknowledge additional support from the NSF under grant No. DMR-1005880. 
\end{acknowledgments}

\end{document}